# The reaction $pp \to pp\pi^0$ near threshold

C. Hanhart, J. Haidenbauer, A. Reuber, C. Schütz and J. Speth

*Institut für Kernphysik, Forschungszentrum Jülich GmbH, D–52425 Jülich, Germany*

**Abstract**

We perform a momentum-space calculation of the reaction $pp \to pp\pi^0$ near threshold. Direct production and pion rescattering are taken into account. For the latter a T-matrix obtained from a microscopic model of the $\pi N$ interaction is employed. The effects of some approximations commonly used in studies of $pp \to pp\pi^0$ are discussed. We find that those approximations overestimate the cross section and also modify its energy dependence. We obtain a significant contribution from rescattering, but not nearly enough to describe the data.

## 1 Introduction

The high precision data of the reaction $pp \to pp\pi^0$ close to pion production threshold from the Indiana University Cyclotron Facility (IUCF) [1,2] continue to attract the attention of theorists. Early theoretical studies [3,4] followed closely the conventional approach to s-wave meson production which goes back to the work of Koltun and Reitan from 1966 [5]. In this model two production mechanisms are considered: (i) direct production of a pion, depicted in Fig. 1a, (ii) production of a pion which first scatters off the other nucleon before emission, as shown in Fig. 1b. It became clear that such a model grossly underestimates the empirical cross section [3,4]. On the other hand, the predicted energy dependence, which in this model is determined essentially by phase-space factors and the $pp$ final-state interaction [1], was found to be in agreement with the IUCF data.

A first quantitative description of the $pp \to pp\pi^0$ cross section was achieved by Lee and Riska [6]. These authors considered in addition effects from meson-exchange currents due to the exchange of heavy (i. e. short ranged) mesons, as shown in Fig. 1c. The resulting contributions (in particular that of the $\sigma$ meson) enhance the cross section near threshold by a factor of 3-5 [6,7] and thus eliminate most of the underprediction found in earlier investigations. As a consequence it was argued in ref. [6] that the reaction $pp \to pp\pi^0$ could provide direct information on the short range part of the $NN$ force.

However, in a very recent investigation by Hernández and Oset [8] an alternative explanation for the missing strength in the $\pi^0$ production close to threshold is presented. Following the idea of Hachenberg and Pirner [9] the off-shell properties of the $\pi N$ amplitude are taken into account explicitly when evaluating the rescattering diagram. Since the off-shell $\pi N$ isoscalar amplitude is, in general, much larger than its on-shell value, which was used in former computations, the contribution from rescattering is now considerably enhanced and the resulting $pp \to pp\pi^0$ cross section can even be brought into agreement with the experimental data. Thus their result suggests that the reaction $pp \to pp\pi^0$ might be a tool to pin down the off-shell behavior of the $\pi N$ interaction.

In the present paper we re-examine critically some of the published results. In these publications several approximations to the pion production vertex and the kinematics have been tacitly taken over from the very first investigation by Koltun and Reitan [5] and we want to shed some light on their validity. It will turn out that these approximations have a non-negligible influence on the magnitude as well as the energy dependence of the $pp \to pp\pi^0$ cross section. We also review the importance of contributions from pion rescattering. We believe that this is necessary because in the work of Hernández and Oset phenomenological prescriptions for the off-shell extrapolation of the $\pi N$ amplitude are applied. In the present paper a realistic meson-theoretical model for the $\pi N$ interaction [10] is employed. Furthermore these authors use some simplifications in the kinematics in evaluating the rescattering diagram which are avoided in the present work. We confirm qualitatively the results reported in ref. [8] but we find rather smaller rescattering contributions.

Our calculations are carried out in momentum space. This allows us to retain all non-localities of the $NN$ potential as well as of the $\pi N$ interaction. Furthermore we can avoid approximations at the pion production vertex. On the other hand we have to deal with the difficulty of incorporating the Coulomb interaction in the momentum-space calculations. It has been shown in the past [11] that the strong (on-shell) T-matrix in the presence of the Coulomb interaction, $T^{cs}$, can be calculated reliably with a method proposed by Vincent and Phatak [12]. However, $T^{cs}$ is needed now half-off-shell where this method is not applicable. In principle one could solve the scattering equation in Coulomb-wave representation [13] which is however a delicate numerical problem. Therefore we use the following prescription for extending $T^{cs}$ off-shell,

$$T^{cs}(q,k) = \frac{C(\gamma_q)}{C(\gamma_k)} \frac{T^s(q,k)}{T^s(k,k)} T^{cs}(k,k) , \qquad (1)$$

where $k$ and $q$ denote the on-shell and off-shell momentum, respectively. $T^s(k,k)$ and $T^s(q,k)$ are the on-shell and half-off-shell T-matrices for the strong interaction alone. The Coulomb penetration factor is given by

$$C^2(\gamma_k) = \frac{2\pi\gamma_k}{e^{2\pi\gamma_k}-1}; \quad \gamma_k = \frac{M}{2}\frac{1}{\alpha k}, \qquad (2)$$

with $M$ being the nucleon mass and $\alpha$ the fine-structure constant. We have checked the reliability of this prescription by applying it to the Reid soft-core potential [14], where the half-off-shell T-matrix including Coulomb interaction can be also obtained from solving the (inhomogeneous) Schrödinger equation [15]. The deviations occurring in the resulting $pp \to pp\pi^0$ cross sections are of the order of 1%.

We consider only the lowest partial waves in the outgoing channel; the $pp$ pair is taken to be in the $^1S_0$ and the pion in an s-wave relative to the nucleon pair. From angular momentum and parity conservation it then follows that the initial $pp$ state has to be $^3P_0$. Distortions in the initial and final $NN$ states are taken into account. The distorted waves are expressed in terms of the (half-off-shell) T-matrices in the standard way and the occurring loop integrals are then computed numerically.

## 2 Direct pion production

In this section we study the direct pion production mechanism shown in Fig. 1a. Here the Reid soft-core potential is employed for the distortions in the initial and final $pp$ states. For this interaction model we can compare our results with the ones obtained by Meyer et al. [2]. The Coulomb interaction is neglected in this section.

Following previous investigations we use pseudo-vector coupling for the $\pi NN$ vertex. This leads to the following structure for the pion production vertex (omitting an additional term cubic in the momenta [16] which is negligible near threshold),

$$M_{fi} \propto \sqrt{\frac{\epsilon_p \epsilon_{p'}}{\omega_q E_p E_{p'}}} \left[\vec{\sigma}\cdot(\vec{p}-\vec{p}') - \omega_q \vec{\sigma}\cdot(\frac{\vec{p}}{\epsilon_p}+\frac{\vec{p}'}{\epsilon_{p'}})\right], \qquad (3)$$

where $\vec{p}$ ($\vec{p}'$) is the incoming (outgoing) nucleon momentum, and $\epsilon_p = E_p + M$ with the nucleon energy $E_p = \sqrt{M^2+\vec{p}^2}$. $\omega_q = \sqrt{m_\pi^2+\vec{q}^2}$ is the energy of the pion with momentum $\vec{q} = \vec{p}-\vec{p}'$.

In previous calculations various approximations have been made. For instance the $\vec{\sigma}\cdot(\vec{p}-\vec{p}\,')$ term is omitted[1], the energies $E_p$, $E_{p'}$, and $\omega_q$ are replaced by the respective masses and simplifications are made in the kinematics of the pion. We want to examine the effects of these approximations on the cross section for $pp \rightarrow pp\pi^0$ step by step. We start with the full vertex (eq.(3)) which leads to the results given by the solid line in Fig. 2.

For the non-relativistic approximation where in eq. (3) $E_p$, $E_{p'}$ and $\omega_q$ are replaced by $M$ and $m_\pi$, respectively, the dotted curve is obtained. Clearly this leads to a 30% increase in the cross section (cf. Fig. 2). Another approximation commonly used concerns the evaluation of the momentum of the produced pion. It is done in the final $NN$ center-of-momentum frame rather than in the $\pi NN$ system. This means that in the kinematical relations the reduced mass of the pion relative to the $NN$ system ($\mu_{(NN)\pi} = \frac{2Mm_\pi}{2M+m_\pi}$) is replaced by the pion mass $m_\pi$. The result is an increase in the allowed maximum pion momentum $q^\pi_{max}$ of about 7 %, which in turn causes an additional enhancement of the total cross section (cf. the difference of the dashed and the dotted curve in Fig. 2). Evidently, the two approximations discussed so far lead to an overestimate of the cross section by almost a factor 2.

Finally, in most previous calculations the first term in eq. (3) has been omitted. This term is proportional to the pion momentum $\vec{q}$ and therefore it has been argued that it should be suppressed in the threshold region where the new IUCF data have been taken. However, our results shown in Fig. 2 demonstrate that this is true only extremely close to threshold — cf. the dot-dashed and the dashed curves. Already for incident energies corresponding to $\eta$ ($= q^\pi_{max}/m_\pi$) values as small as 0.2 the first term has a noticeable influence on the cross section and its influence increases dramatically with increasing energy. Evidently, the opposite signs of the two terms in eq. (3) cause a destructive interference leading to a decrease in the cross section.

Fig. 2 contains also the result of a calculation by Meyer et al. [2] which is based on all the approximations discussed. It can be seen that our momentum space calculation agrees very well with their configuration space result once we have incorporated those approximations.

---

[1] This applies only to the direct production diagram. For the pion production vertex in the rescattering diagram this term is kept whereas the second term in eq. (3) is usually omitted.

## 3 Pion rescattering

The second pion production mechanism we want to study is pion rescattering (Fig. 1b). In most calculations the $\pi N$ scattering amplitude has been derived from a phenomenological effective Hamiltonian [5]

$$H = 4\pi \frac{\lambda_1}{m_\pi} \bar{\Psi} \vec{\phi} \cdot \vec{\phi} \Psi + 4\pi \frac{\lambda_2}{m_\pi^2} \bar{\Psi} \vec{\tau} \Psi \cdot \vec{\phi} \times \partial_0 \vec{\phi} \qquad (4)$$

where $\lambda_1$, $\lambda_2$ are fixed by the (empirical) $S_{11}$ and $S_{31}$ pion nucleon scattering lengths [6]. The isovector term (proportional to $\lambda_2$) does not contribute to the rescattering diagram in fig. 1b. As $\lambda_1$ is very small ($\lambda_1 = 0.005$ according to Höhler et al. [17]; $\lambda_1 = -0.0013$ following Arndt et al. [18]) it has usually been found that the rescattering contribution to the $pp \to pp\pi^0$ cross section obtained from the effective Hamiltonian in eq. (4) is small.

However, it is well known that the smallness of the isoscalar $\pi N$ on-shell amplitude is due to a strong cancellation between different isospin amplitudes. The situation is rather different once one moves off-shell — and it is indeed the off-shell $\pi N$ amplitude which enters into the rescattering diagram (Fig. 1b). This fact is exploited in a recent investigation by Hernández and Oset [8]. Using two different phenomenological off-shell extrapolations these authors demonstrate that the pion-rescattering mechanism can enhance the $pp \to pp\pi^0$ cross section by a factor of 5 or more.

In the following we re-examine the findings of Hernández and Oset. In our investigation however, a microscopic meson–exchange model of $\pi N$ interaction is implemented which has recently been constructed in Jülich. This model includes the conventional (direct and crossed) pole diagrams involving the nucleon and the $\Delta$-isobar; the meson exchanges in the scalar ($\sigma$) and vector ($\rho$) channels are derived from correlated $2\pi$ exchange. Details can be found in ref. [10]. Here we use a slightly modified version where the form factors are energy–independent and the antibaryon contributions have been left out. These modifications are made to allow an extrapolation of the model to negative energies as required in the present three-particle context. After readjustment of its free parameters this model yields a good description of low–energy $\pi N$ scattering, comparable to the results shown in ref. [10]. The resulting $s$–wave scattering lengths are $a_1 = 0.173 m_\pi^{-1}$ and $a_3 = -0.084 m_\pi^{-1}$ leading to a value of $\lambda_1 = -0.001$ which is in line with the value given in ref. [7]. The half-off-shell T-matrix produced by this model is shown in Fig. 3 at $\pi N$ threshold.

Besides of using a microscopic $\pi N$ model there are also some differences in the evaluation of the rescattering contribution. In Ref. [8] only the first term of the pion production vertex in eq. (3) is used whereas we use again the full vertex. Furthermore the four-momentum of the exchanged pion, $q$, is fixed by $q^0 = 70\ MeV$ and $|\vec{q}| = 370\ MeV/c$, corresponding to the particular kinematical situation at the pion production threshold when there are no distortions in the initial and final $pp$ states. As a consequence only the half-off-mass-shell isoscalar $\pi N$ amplitude enters into the calculation of Hernández and Oset (cf. Ref. [8]). We allow full freedom in the momentum transferred by the pion as it appears in the loop integrations when the initial- and final-state interactions are taken into account. In this case the $\pi N$ T-matrix is needed fully off-shell. It depends on the relative momenta of the $\pi N$ states and on the energy available in the $\pi N$ subsystem.

The contribution of the rescattering mechanism to the $pp \to pp\pi^0$ cross section is illustrated in Fig. 4. Note that now the Bonn potential OBEPF [19] is used for the initial and final $pp$ interaction throughout and the Coulomb interaction is taken into account. For the pion production vertex we use the same parameters as for the $\pi NN$ vertex in the $NN$ potential, namely the value $f^2/4\pi = 0.0795$ for the pion-nucleon coupling constant and a monopole form factor with cutoff mass $\Lambda_\pi = 1.75\ GeV$. It can be seen that rescattering increases the cross section by a factor of around 3.5 compared to the rate for direct production. Our result therefore confirms qualitatively the findings reported in ref. [8]. However, the effect of rescattering with our microscopic model for the $\pi N$ T-matrix is obviously smaller. This can be understood if we contrast the off-shell behavior of the $\pi N$ T-matrices used in the two calculations, as is done in Fig. 3. We want to point out that in our calculation the rescattering contribution enhances the cross section even though $\lambda_1$ is negative; the effective Hamiltonian of eq. (4) leads to a decrease in the cross section for negative $\lambda_1$ [7].

As can be seen in Fig. 4 our predictions underestimate the empirical data [2,20] by a factor of 3.7. Scaling the results by this factor (dot-dashed curve in Fig. 4) reveals that also the energy dependence of the data is not reproduced. The latter feature is in contrast to essentially all previous calculations. It is caused by the additional $\eta$-dependence introduced by the first term of the pion production vertex (eq. (3)). Thus the shape of the total cross section as a function of the energy is no longer determined by phase-space factors and the final-state interaction alone. For the sake of clarity we have repeated the calculation using only the second term of eq. (3) for the direct production and only the first term of the pion-production vertex for the rescattering contribution, as it is usually done in the older calculations. The result, scaled by a factor of 1.8, is also shown in Fig. 4 (dotted curve).

Coming back to the direct production diagram oncemore we want to note that Niskanen also kept the first term in eq. (3) in his investigation [4]. He observes a similar interference effect as we do but in his case this interference occurs at somewhat higher $\eta$ values so that he can still describe the energy dependence of the IUCF data. We believe that this is due to the following: Niskanen uses the phenomenological Hamiltonian of eq. (4) but he allows a dependence of $\lambda_1$ on the (on-shell) $\pi N$ momentum. Since $\lambda_1$ increases rather rapidly with increasing energy (cf. Fig. 3 in Ref. [21]) the s-wave $\pi N$ rescattering contribution compensates, to a large extent, the decrease in the cross section introduced by the first term of the direct pion production in Niskanen's work [21]. On the other hand, in our calculation based on the full off-shell dependence of the $\pi N$ amplitude the main contributions from rescattering come from off-shell momenta of typically 200-400 $MeV/c$. For these off-shell momenta the $\pi N$ isoscalar s-wave amplitude depends only weakly on the energy and therefore no such compensation occurs.

## 4 Discussion and conclusion

Our investigations reveal that several of the approximations applied in previous studies of the reaction $pp \to pp\pi^0$ close to pion production threshold are not justified. The correct treatment of the pion production vertex reduces the $\pi^0$ production rate by a factor of 2. However, even more serious is the modification in the energy dependence of the $pp \to pp\pi^0$ cross section. Whereas the calculations using the approximated vertex can reproduce the energy dependence of the IUCF data, the result based on the full vertex starts to deviate already at $\eta \approx 0.3$ (cf. Fig. 4). This cannot be compensated by, for example, taking into account $p$-waves in the $NN$- and $\pi N$-sector since their contributions do not set in before values of $\eta$ around 0.5 (as we found in test calculations). Therefore it seems that a genuinely new mechanism is needed. A possible solution could be corrections from heavy-meson exchange. Indeed the figures shown in refs. [6,7] suggest that the contributions from heavy-meson exchange modify the $\eta$ dependence of the cross section to the right direction. However, it is unlikely that these modifications will be sufficient to remedy the discrepancy observed in our calculation. Another possibility which was been advocated in the literature [22,23] is the coupling between the final $pp\pi^0$, $np\pi^+$, and $d\pi^+$ channels. In this context we would like to remark that the threshold of the $np\pi^+$ channel lies close to $\eta = 0.3$.

The rescattering diagram gives rise to an appreciable enhancement of the $pp \to pp\pi^0$ cross section provided that the off-shell behavior of the $\pi N$ amplitude is properly accounted for. With our microscopic $\pi N$ model we obtain an enhancement factor of 3.5 which more than compensates the reduction which we get from the correct treatment of the vertex in the direct production diagram. However, we are still far away from the data. Moreover, this compensation is true only up to $\eta \approx 0.3$. For larger $\eta$ values the correct operator structure of the pion production vertex introduces an energy dependence which is not completely compensated by the rescattering contribution.

The off-shell behavior of the $\pi N$ amplitude used in the present calculation is completely determined by the dynamical ingredients of the $\pi N$ model and the fit to the $\pi N$ data. However, in the original models presented in ref. [10], which differs in its dynamical ingredients (inclusion of antibaryon contributions), the maximum of the $\pi N$ amplitude (in Fig. 3) would be roughly twice as high and accordingly a larger contribution from the rescattering mechanism can be expected. We have left out those antibaryon contributions for consistency with the $NN$ model used and to permit an extrapolation to negative energies. Given the uncertainties in the off-shell properties of the $\pi N$ amplitude it is difficult at present to obtain a reliable estimate of the contribution from rescattering. Clearly a consistent description is needed where the $NN$ and the $\pi N$ systems are treated on the same footing if one wants to make real progress. We will continue working in this direction.

## Acknowledgments

We would like to thank H. Haberzettl, K. Holinde, M. Macfarlane and F. Osterfeld for useful discussions and critical comments.

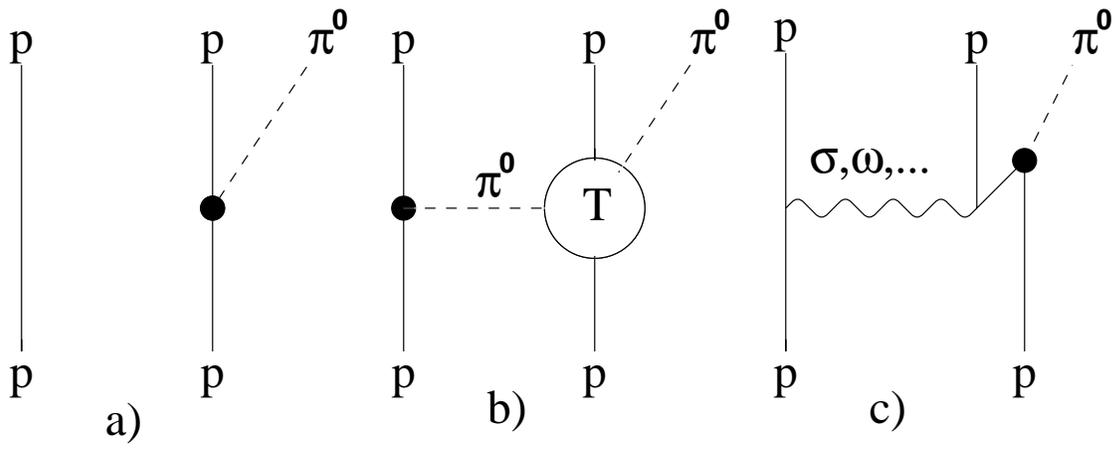

Fig. 1. Mechanisms for pion production: (a) direct production, (b) pion rescattering, (c) production incorporating heavy-meson exchange.

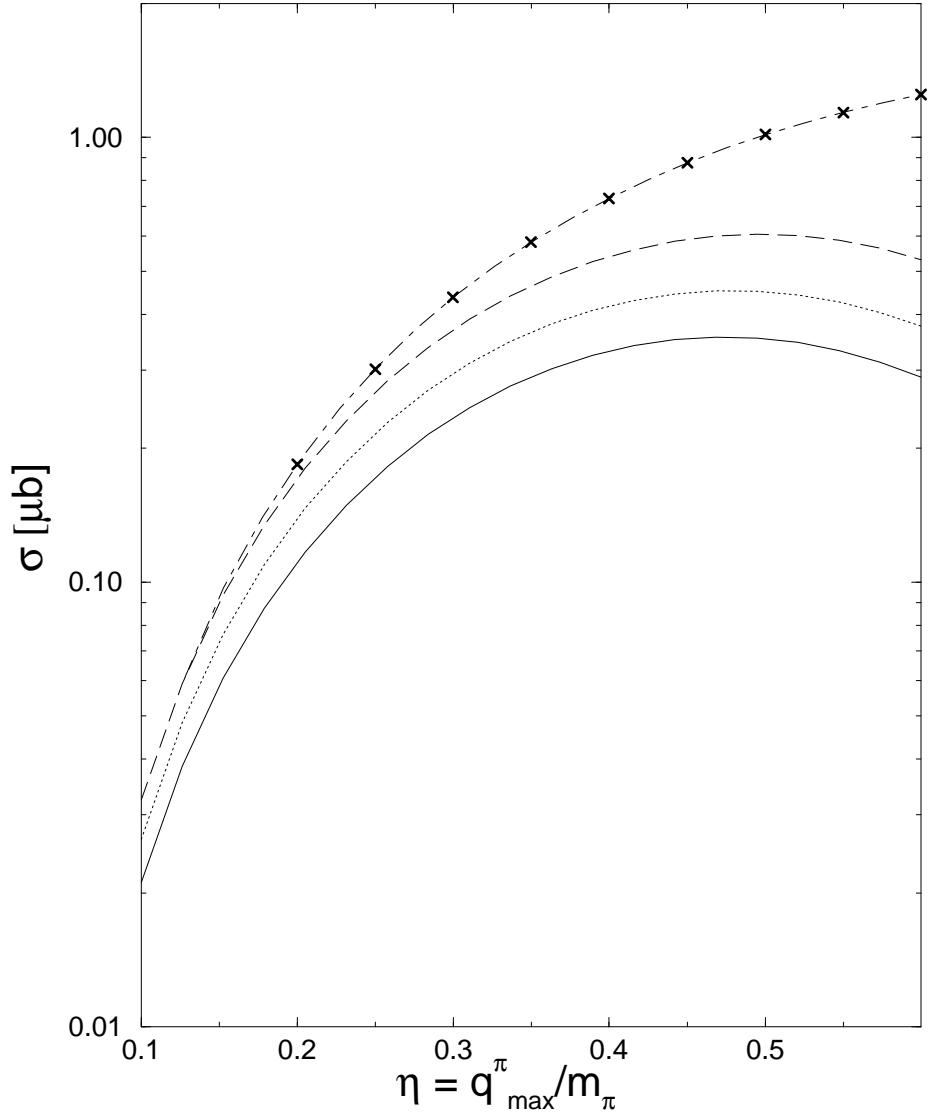

Fig. 2. Effects of different approximations at the pion production vertex on the $pp \to pp\pi^0$ cross section. The Reid soft-core potential is used for the initial- and final-state interaction of the two protons. The solid line is the result for the full vertex (eq.(3)). The dotted curve is obtained when the energies of the particles are replaced by their masses. For the dashed line in addition the reduced pion mass is replaced by $m_\pi$. Finally, omitting the first term in eq. (3) yields the dot-dashed curve. This approximation corresponds to the results obtained by Meyer et al. [2], which are indicated with crosses.

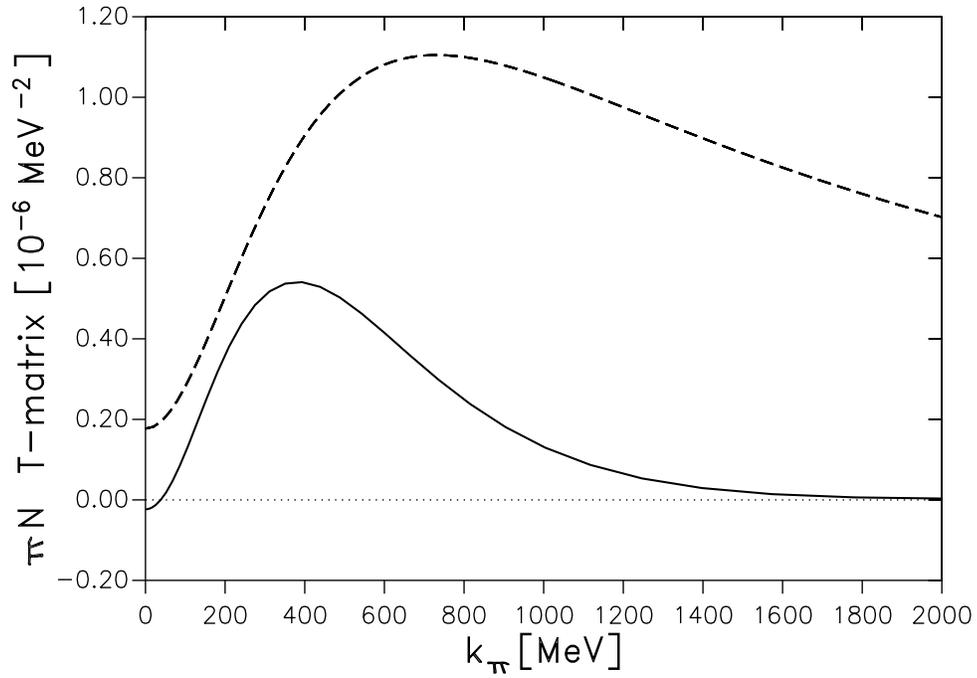

Fig. 3. Half-off-shell behavior of the $\pi N$ T-matrix at $\pi N$ threshold. The solid line denotes the model used in the present calculation. The dashed curve is the model by Hamilton [24] used in Ref. [8].

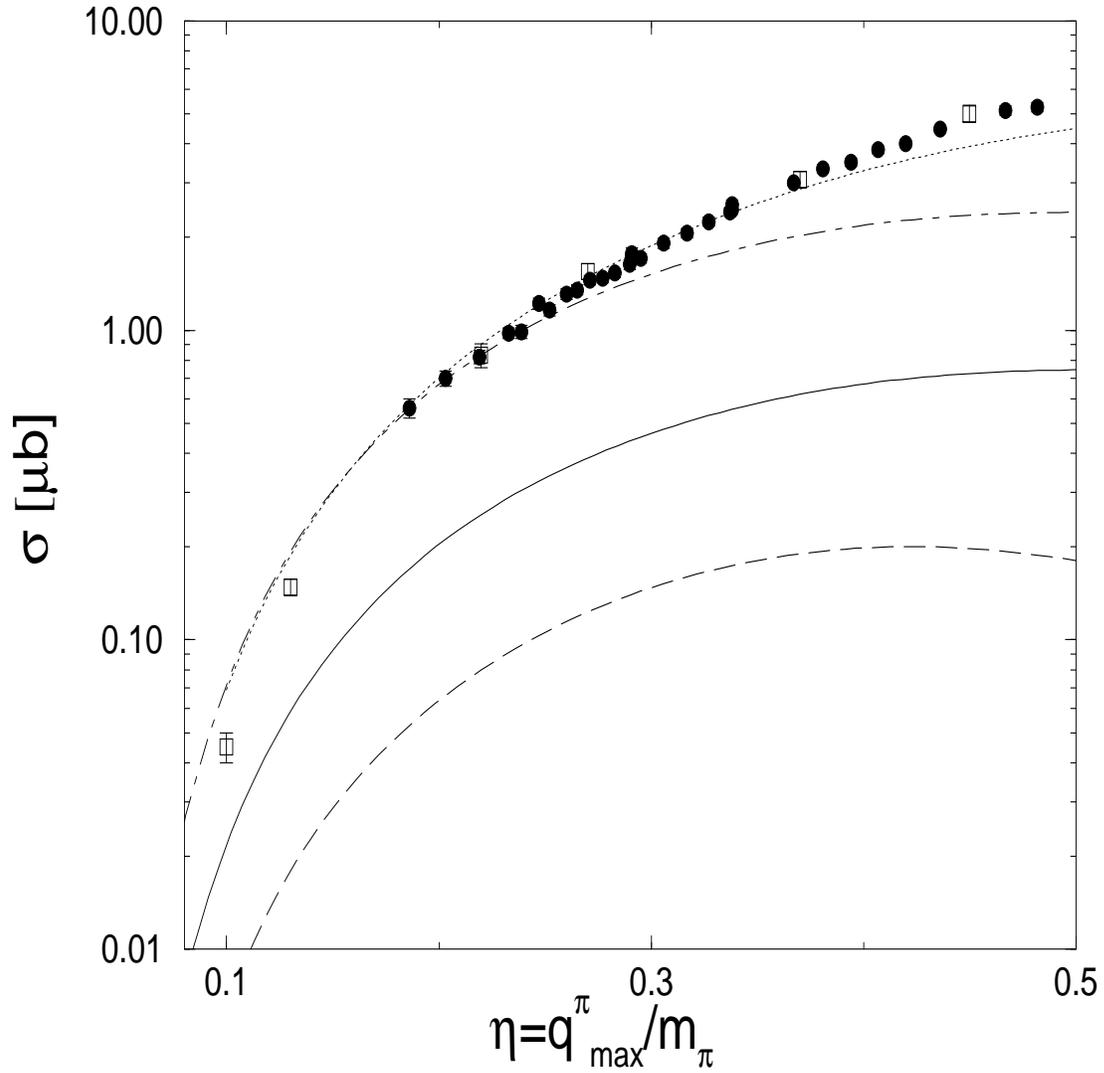

Fig. 4. Results for the reaction $pp \to pp\pi^0$ using the model OBEPF [19] for the initial- and final-state interaction of the two protons. The solid curve represents our full calculation whereas the dashed line is obtained when only the direct production is taken into account. The dot-dashed curve is the full result scaled by a factor of 3.7. The dotted curve is the full calculation based on the approximative treatment of the pion production vertex (cf. text). It is scaled by a factor of 1.8. The data points are taken from refs. [2] (dots) and [20] (squares).